\documentstyle[preprint,eqsecnum,aps,epsfig]{revtex}
\begin{document}
\draft
\preprint{APCTP-1999002, SOGANG-HEP-251/99}
\title{Maximum Mass of Boson Stars Formed by Self-Interacting Scalar Fields} 
\author{Jeongwon Ho\footnote{E-mail address : 
jwho@apctp.kaist.ac.kr}}
\address{
Asia Pacific Center for Theoretical Physics,\\
207-43 Cheongryangri-dong, Dongdaemun-gu, Seoul 103-102,
Korea}
\author{Sungjoon Kim\footnote{E-mail address : 
sjkim@physics3.sogang.ac.kr} and 
Bum-Hoon Lee\footnote{E-mail address : 
bhl@ccs.sogang.ac.kr}
}
\address{
Department of Physics and Basic Science 
Research Institute,\\
Sogang University, C.P.O.Box 1142, Seoul 
100-611, Korea}
\date{February 1999}
\maketitle

\begin{abstract}
We make {\it analytic} derivation for maximum masses of stable boson stars
formed by scalar fields with {\it any} higher order self-interactions
and show that those are equivalent to numerical results.
It is shown that the contribution of the higher order
self-interaction terms to the maximum mass decreases as $(m/M_p)^2$
power.
\end{abstract}
\

\narrowtext

\section{Introduction}
Though there is no evidence for the presence of a fundamental
scalar field, it has played an important role for the study of the
earliest stages of the universe \cite{kolb}\cite{linde}.
According to this interest, the
boson star, which was first discovered theoretically by Kaup \cite{kaup} and
by Ruffini and Bonazzola \cite{ruffini}, has
become an interesting subject in its
own right as well as in the problem of the cosmological missing mass
problem. This configuration has been studied in various theories
\cite{colpi}\cite{bij}\cite{jetzer}\cite{henriques}.
Their stability \cite{gleiser} \cite{kusmartsev}
and structure \cite{peebles} have
been also investigated in some literatures.
(for reviewing the boson star, see Refs.\cite{liddle} and \cite{mielke},
and other references are cited therein). 

Boson stars are formed by soliton-type configurations which are held
together simply by their self-generated gravitational field and are
only prevented from gravitational collapse by the Heisenberg uncertainty
principle. Thus, we can easily calculate the order of radius $R_0$, maximum
mass $M^{max}_0$, and energy density $\rho_0 $ of a spherically symmetric
stable boson star \cite{kaup}\cite{ruffini} as
\begin{equation}
\label{bare}
R_0 \sim 1/m,~~M^{max}_0 \sim M_p^2/m,~~\rho_0 \sim M_p^2m^2,
\end{equation}
where $m$ and $M_p$ are the boson mass and the Planck mass, respectively
(we use the unit set by $\hbar = c =1 $). Then,
the magnitude of the central value of the scalar field leading to the most
massive stable boson stars has the order of the Planck mass,
$|\phi | \sim M_p $. In Eq.(\ref{bare}), we see that the maximum mass
of a boson
star is much less than the Chandrasekhar mass, $ M_{ch} \sim
M_p^3/m^2 $. For a bosonic particle with mass 100 GeV, comparing with a
neutron star, the maximum mass, radius, and density of a boson
star are $10^{9}kg \sim 10^{-21}M_{neutron} $, $10^{-18} m \sim
10^{-22}R_{neutron}$, and $10^{63}kg/m^3 \sim 10^{51} \rho_{neutron}$,
respectively \cite{mielke}.

The maximum mass of such a stable ``mini-boson star'' is too small to be
a candidate for the missing mass. It is worthy of notice that
why a boson star has a much smaller mass than a typical
fermionic star is due to the fact that a boson star is only prevented
from gravitational collapse by the Heisenberg uncertainty principle instead
of the Pauli exclusion principle. In other words, the Heisenberg uncertainty
principle is characterized by a much smaller length scale than
that of the Pauli exclusion principle. It has been enhanced by Colpi, Shapiro
and Wasserman \cite{colpi}, who considered the boson star formed with
self-interacting scalar fields. They have shown that when the order
of the coupling constant $\lambda $ is about unity, the boson star may have
the maximum mass comparable to that of a fermionic star.
\begin{equation}
\label{masslam}
M^{max}_{\lambda} \sim 0.22\Lambda^{1/2}M_p^2/m 
\sim 0.22\lambda^{1/2} M_{ch},
\end{equation}
where $\Lambda $ is a dimensionless quantity defined by
$\Lambda \equiv \lambda M_p^2/4\pi m^2 $. The origin of this drastic change
is that the introduced self-interaction term in matter Lagrangian generates
a repulsive force. Thus, the characteristic length scale of the theory
becomes large, and the boson star becomes large and massive.

In this paper, we shall make {\it analytic} derivation for the maximum mass
in Eq.(\ref{masslam}) calculated by numerical analysis. (Analytic
derivation of maximum mass of boson star was also appeared in
Ref.\cite{tkachev}.) Our analytic
derivation will be also applied to the boson star formed by scalar fields
with {\it any} higher order self-interactions. Then, a general form of
the maximum mass including the self-interaction effect is obtained
and it will be shown that the contribution of the higher order
self-interaction terms to the maximum mass decreases as $(m/M_p)^2$
power. In order to confirm our analytic calculations,
we shall make numerical analysis for the scalar field with the quartic
and the sixth order self-interactions\footnote{For the boson star
formed by the scalar field with the sixth order self-interaction,
see Refs.\cite{mielke2} and \cite{kusmartsev}}.
In Ref.\cite{mielke}, the authors
chose the magnitude of the central value of the scalar field
to be the order of the Planck mass, $|\phi | \sim M_p $, even for the
self-interacting scalar fields. However, we shall show that when
the contribution of the self-interaction becomes important, it has
the order of the mass of the scalar field, $|\phi| \sim  m \lambda^{-1/2}$
(This relation seems to be singular in the limit of $\lambda \rightarrow 0$.
But, our derivation of the relation will be given only in the case
that the limit is invalid.) This is consistent with the numerical
calculation given in Ref.\cite{colpi}.

This paper is organized as follows;
In Sect.II, the order of the maximum mass of a stable boson star formed by
the scalar field with the quartic and the sixth order self-interactions
is analytically derived. Then, the formulation is generalized to the scalar
field with any higher order self-interactions. Our derivation is confirmed
in numerical analysis given in Sect.III.
In Sect.IV, we briefly summarize and discuss our results.

\section{Analytic Derivation for Maximum Masses}
In this section we make analytic derivations of the maximum masses of
boson stars formed by self-interacting scalar fields. Firstly,
this is done for the cases that the matter Lagrangian includes the quartic
and the sixth order self-interaction terms. Then, we are going to extend
our formulation for any higher order self-interacting bosons.

We begin with the Lagrangian of the scalar field $\phi $ minimally
coupled to gravity given by
\begin{equation}
\label{lagrangian}
{\cal L}=-{1\over2}g^{\mu\nu}\phi_{;\mu}^*\phi_{;\nu}-{1\over2}m^2
{|\phi|}^2-{1\over4}\lambda{|\phi|}^4-{1\over6}\gamma{|\phi|}^6,
\end{equation}
where $\lambda$ and $\gamma $ are the coupling constants.

As mentioned above, when one is not concerned about the self-interaction,
the magnitude of the central value of the scalar field leading to the most
massive stable boson stars $|\phi |$ is the order of the Planck mass.
When the contribution of the self-interaction becomes important,
putting $|\phi |$ is of the order of the Planck mass, however, will give rise
to some inconsistencies. This is because the self-interaction generates a
repulsive force and the characteristic length scale of the theory becomes
larger. To see this, consider
only the quartic self-interaction term ($\gamma =0$) in the Lagrangian
(\ref{lagrangian}). If one set the order of $|\phi |$ with that of the
Planck mass, the energy density including the quartic self-interaction term
$\rho_{\lambda} $ (Eq.(14) in Ref.\cite{mielke}) becomes
\begin{equation}
\label{densitylamf}
\rho_{\lambda} \sim m^2 {|\phi|}^2 + \lambda{|\phi|}^4 
\sim m^2 M_p^2 (1+ \Lambda ).
\end{equation}
Comparing Eqs.(\ref{bare}) and (\ref{densitylamf}),
we see that this corresponds to the energy density of a star formed from
non-interacting bosons with rescaled mass $m \rightarrow m(1
+ \Lambda )^{1/2} $ and the order of the radius of the boson star is given by
\begin{equation}
\label{radiuslamf}
R_{\lambda} \sim (1+ \Lambda )^{-1/2}/m.
\end{equation}
Thus, the maximum mass of the boson star becomes
\begin{equation}
\label{masslamf}
M^{max}_{\lambda} \sim \rho_{\lambda} R_{\lambda}^3
\sim (1+ \Lambda )^{-1/2} M_p^2/m
\sim \Lambda^{-1/2} M_p^2/m.
\end{equation}
Note that the maximum mass in (\ref{masslamf}) is not equivalent to the
numerical result (\ref{masslam}) given in Ref.\cite{colpi}.
Moreover, $\rho_{\lambda} $ in Eq.(\ref{densitylamf})
is much greater than $\rho_0$ in Eq.(\ref{bare}) and
$R_{\lambda}$ in Eq.(\ref{radiuslamf}) much less than $R_0$ in
Eq.(\ref{bare}). It seems to be inconsistent with the fact that
the introduced self-interacting potential generates a repulsive force.

Now, turn to the energy density including the quartic self-interaction term.
Since we expect from the field equation that the contribution of the
self-interaction term is comparable to the mass term in the energy density,
$m^2 |\phi |^2 \sim \lambda |\phi |^4 $, the order of $|\phi |$ becomes
\begin{equation}
\label{potenergylam}
\frac{\lambda |\phi |^4}{m^2 |\phi |^2} \sim \Lambda \frac{|\phi |^2}{M_p^2}
\sim {\cal O}(1),~~|\phi | \sim M_p \Lambda^{-1/2} \sim m \lambda^{-1/2}.
\end{equation}
Note that $|\phi|$ has the order of the scalar mass not the Plank mass.
In fact, we can see that in Fig.1 of Ref.\cite{colpi}, the central value of
$|\phi |$ decreases with increasing $\Lambda $. (The Fig.1 of
Ref.\cite{colpi} was plotted only up to the value $\Lambda =300$. However,
the dimensionless constant $\Lambda$ is very large, $\Lambda \approx
10^{34}\sim 10^{38}$. For this value, the central density $\sigma_0$ in the
figure would be very small.) On the other hand, the Eq.(\ref{potenergylam})
seems to be singular in the limit of $\lambda \rightarrow 0 $. However,
as mentioned above, the Eq.(\ref{potenergylam}) is only available
in the case that the self-interaction term is so large to be comparable
to the mass term in the energy density. The energy density becomes
\begin{equation}
\label{densitylam}
\rho_{\lambda} \sim m^2 {|\phi|}^2 + \lambda{|\phi|}^4 
\sim m^2M_p^2 \Lambda^{-1}.
\end{equation}
Again, this corresponds to the energy density of a star formed from
non-interacting bosons with rescaled mass $m \rightarrow m \Lambda^{-1/2}$.
Thus, the radius of the boson star becomes
\begin{equation}
\label{radiuslam}
R_{\lambda} \sim \Lambda^{1/2}/m.
\end{equation}
In Eqs.(\ref{densitylam}) and (\ref{radiuslam}),
the density and the radius become dilute and large, respectively,
and it is consistent with the fact that a repulsive force
is involved. Finally, using Eqs.(\ref{densitylam}) and (\ref{radiuslam}), we
obtain the maximum mass of the boson star
equivalent to the numerical result in (\ref{masslam}) up to constant
\begin{equation}
\label{masslambda}
M^{max}_{\lambda} \sim \Lambda^{1/2}M_p^2/m \sim \lambda^{1/2} M_{ch}.
\end{equation}

Let us now consider the sixth order self-interacting bosons.
Following the above formulation, the central value of the scalar field
is given by
\begin{equation}
\label{potenergylamgam}
\frac{\lambda |\phi |^4 + \gamma |\phi |^6}{m^2 |\phi |^2}
\sim \left(\frac{|\phi |}{M_p} \right)^4\left[
\Lambda \left(\frac{M_p}{|\phi |} \right)^2 +\Gamma \right]
\sim \left(\frac{|\phi |}{M_p} \right)^4
(\Lambda^2 +\Gamma ) \sim {\cal O}(1),
\end{equation}
\begin{equation}
\label{phival}
|\phi | \sim M_p (\Lambda^2 + \Gamma)^{-1/4},
\end{equation}
where $ \Gamma\equiv \gamma M_p^4/(4\pi m)^2$.
Note that we used the mean
field approximation on the second step in Eq.(\ref{potenergylamgam}), i.e.,
the relation $|\phi | \sim M_p \Lambda^{-1/2} $ given in
Eq.(\ref{potenergylam}).
Thus, the energy density and radius of the boson star are given by
\begin{equation}
\label{densitylamgam}
\rho_{\lambda \gamma} \sim m^2 {|\phi|}^2 + \lambda{|\phi|}^4
+ \gamma |\phi |^6 \sim m^2M_p^2 (\Lambda^2 + \Gamma )^{-1/2},
\end{equation}
\begin{equation}
\label{radiuslamgam}
R_{\lambda \gamma} \sim (\Lambda^2 + \Gamma )^{1/4}/m,
\end{equation}
respectively.
As a result, the maximum mass becomes
\begin{equation}
\label{masslamgam}
M^{max}_{\lambda \gamma} \sim (\Lambda^2 + \Gamma )^{1/4} M_p^2/m
\sim \left(\lambda^2 + \tilde{\gamma}\left(\frac{m}{M_p}\right)^2
\right)^{1/4} M_p^3/m^2
\sim \left(\lambda^2 + \tilde{\gamma} \left(\frac{m}{M_p}\right)^2
\right)^{1/4} M_{ch},
\end{equation}
where $\tilde{\gamma} \equiv \Gamma (4\pi m)^2/M_p^2 = \gamma M_P^2$.
If $\tilde{\gamma} =0$, the Eq.(\ref{masslamgam}) becomes Eq.(\ref{masslam}).
Note that from the field theoretic point of view, $\tilde{\gamma}$ is
about unity and the contribution of the sixth order self-interaction
to the maximum mass is minor as the order of the inverse square Plank mass
comparing with the quartic self-interaction term.

Our above analytic derivations can be easily extended to the case that
the matter Lagrangian involves any higher order self-interaction terms.
The self-interaction terms
can be expressed in terms of dimensionless parameters $\lambda_{(2k + 2)}$
as\footnote{The higher order self-interaction potentials arise from effective
theories is discussed in Ref.\cite{benitez}.}
\begin{equation}
\label{higherterms}
\lambda_{(4)} |\phi|^4 + \frac{\lambda_{(6)}}{M_p^2} |\phi|^6
+ \frac{\lambda_{(8)}}{M_p^4}|\phi |^8 + \cdot \cdot \cdot +
\frac{\lambda_{(2n+2)}}{M_p^{2n-2}}|\phi |^{2n+2}.
\end{equation}
These higher order nonrenormalizable terms are naturally expected from the
gravity as low energy effective theory and the dimensionful constants are given
as the power of $\lambda_{(2k + 2)}/M^k_p$.
Then, the magnitude of the central value of the scalar field becomes
\begin{equation}
\label{magscalarhigher}
|\phi |_{(2n+2)} \sim m \left[
\Sigma_{k=1}^{n} \lambda_{(2k+2)} \left( \frac{m}{|\phi|_{(2n)}} 
\right)^{2(n-k)}\left(\frac{m}{M_p} \right)^{2k-2} \right]^{-1/2n},
\end{equation}
where $|\phi|_{(k)} $ is the magnitude of the scalar field considering
the order of the self-interaction to be $k$. The radius and the
maximum mass are given by
\begin{equation}
\label{radiushigher}
R_{(2n+2)} \sim \frac{M_p}{m^2}
\left[
\Sigma_{k=1}^{n} \lambda_{(2k+2)} \left( \frac{m}{|\phi|_{(2n)}} 
\right)^{2(n-k)}\left(\frac{m}{M_p} \right)^{2k-2} \right]^{1/2n},
\end{equation}
\begin{equation}
\label{masshigher}
M_{(2n+2)} \sim 
\left[
\Sigma_{k=1}^{n} \lambda_{(2k+2)} \left( \frac{m}{|\phi|_{(2n)}} 
\right)^{2(n-k)}\left(\frac{m}{M_p} \right)^{2k-2} \right]^{1/2n}
M_{ch},
\end{equation}
respectively. The Eq.(\ref{masshigher}) can be rewritten by
\begin{eqnarray}
\label{masshigherfor}
M_{(2n+2)} &\sim & \displaystyle
\left[ \lambda^n_{(4)} + \lambda^{n/2}_{(6)} \left(\frac{m}{M_p}\right)^2
+ \lambda^{n/3}_{(8)}\left(\frac{m}{M_p}\right)^4 \displaystyle \right.
\nonumber \\
&& \displaystyle \left.
+ \cdot \cdot \cdot + \lambda_{(2n+2)}\left(\frac{m}{M_p}\right)^{(2n-2)}
 + ~{\rm cross~terms}\displaystyle \right]^{1/2n}
M_{ch}.
\end{eqnarray}
As a result,
in Eq.(\ref{masshigherfor}), the contribution of the higher order
self-interaction terms to the maximum mass decreases as $(m/M_p)^2$
power.
 
\section{Numerical Analysis for Maximum Masses}
In this section, we make numerical analysis to conform maximum masses of
stable boson stars calculated in previous section. We only consider up to the
sixth order self-interaction term. The equations of motion generated from
the Lagrangian (\ref{lagrangian}) are given by
\begin{equation}
\label{eins}
G^\mu_\nu=8\pi GT^\mu_\nu,
\end{equation}
\begin{equation}
\label{scalar}
\nabla^\mu\nabla_\mu\phi-m^2\phi-\lambda{|\phi|}^2\phi-
\gamma{|\phi|}^4\phi=0,
\end{equation}
where the energy-momentum tensor $T_{\mu \nu}$ is given by
\begin{equation}
\label{emt} T^\mu_\nu={1\over2}g^{\mu\nu}\left(\phi^*_{;\sigma}\phi_{;\nu}+
\phi_{;\sigma}\phi^*_{;\nu}\right )-{1\over2}\delta^\mu_\nu
\left[g^{\lambda\sigma}
\phi^*_{;\lambda}\phi_{;\sigma}+m^2{|\phi|}^2+{1\over2}\lambda{|\phi|}^4
+{1\over3}\gamma{|\phi|}^6\right].
\end{equation}
We seek a spherically symmetric, time-independent solution represented
by the line element in Schwarzschild coordinates
\begin{equation}
\label{metric}
ds^2=-B\left(r\right )dt^2+A\left(r\right )dr^2+r^2d\Omega.
\end{equation}
Then, we may choose the ansatz for the scalar
field $\phi $ as
\begin{equation}
\label{ansatz}
\phi\left(r,t\right )=\Phi\left(r\right )e^{-i\omega t},
\end{equation}
and the Eqs.(\ref{eins}) and (\ref{scalar}) are written by
\begin{equation}
\label{eqmo1}
{A^\prime\over A^2x}+{1\over x^2}\left(1-{1\over A}\right )
=\left({\Omega^2\over
  B}+1\right )\sigma^2+{\Lambda\over2}\sigma^4+{\Gamma\over3}\sigma^6
+{\left(\sigma^\prime\right )^2\over A},\end{equation}
\begin{equation}
\label{eqmo2}
{B^\prime\over ABx}-{1\over x^2}\left(1-{1\over A}\right )
=\left({\Omega^2\over
  B}-1\right )\sigma^2-{\Lambda\over2}\sigma^4-{\Gamma\over3}\sigma^6
+{\left(\sigma^\prime\right )^2\over A},\end{equation}
\begin{equation}
\label{eqmo3}
\sigma^{\prime\prime}+\left({2\over x}+{B^\prime\over 2B}-{A^\prime\over
  2A}\right )\sigma^\prime+A\left[\left({\Omega^2\over
  B}-1\right )\sigma-\Lambda\sigma^3-\Gamma\sigma^5\right]=0,
\end{equation}
where prime denotes $d/dx$, and $x\equiv mr$, $\sigma\equiv \left(4\pi G
\right )^{1\over2}\Phi$, and
$\Omega\equiv{\omega/ m}$. If we write 
\begin{equation}
\label{aform}
A\left(x\right )=\left[1-{2{\cal M}\left(x\right )\over x}\right]^{-1},
\end{equation}
the Eq.(\ref{eqmo1}) can be rewritten by
\begin{equation}
  \label{eqmo11}
  {\cal M}^{\prime}=x^2\left[{1\over2}\left({\Omega^2\over
  B}+1\right )\sigma^2+{\Lambda\over 4}\sigma^4+{\Gamma\over
  6}\sigma^6+{\left(\sigma^\prime\right )^2\over {2A}}\right].
\end{equation}
The total mass of a boson star is given by
\begin{equation}
\label{totmass}
M_{\lambda \gamma}={\cal M}\left(\infty\right )\left(M^2_{p}/m\right ).
\end{equation}
We require that boundary conditions be $A\left(0\right )=1$,
$\sigma\left(0\right )=\sigma_0$, ${\sigma^\prime}\left(0\right )=0$
and $\sigma\left(\infty\right )=0$ so that our solution becomes a nonsingular
and asymptotically flat one with finite mass.

\begin{figure}[htbp]
  \begin{center}
    \leavevmode
    \epsfig{file=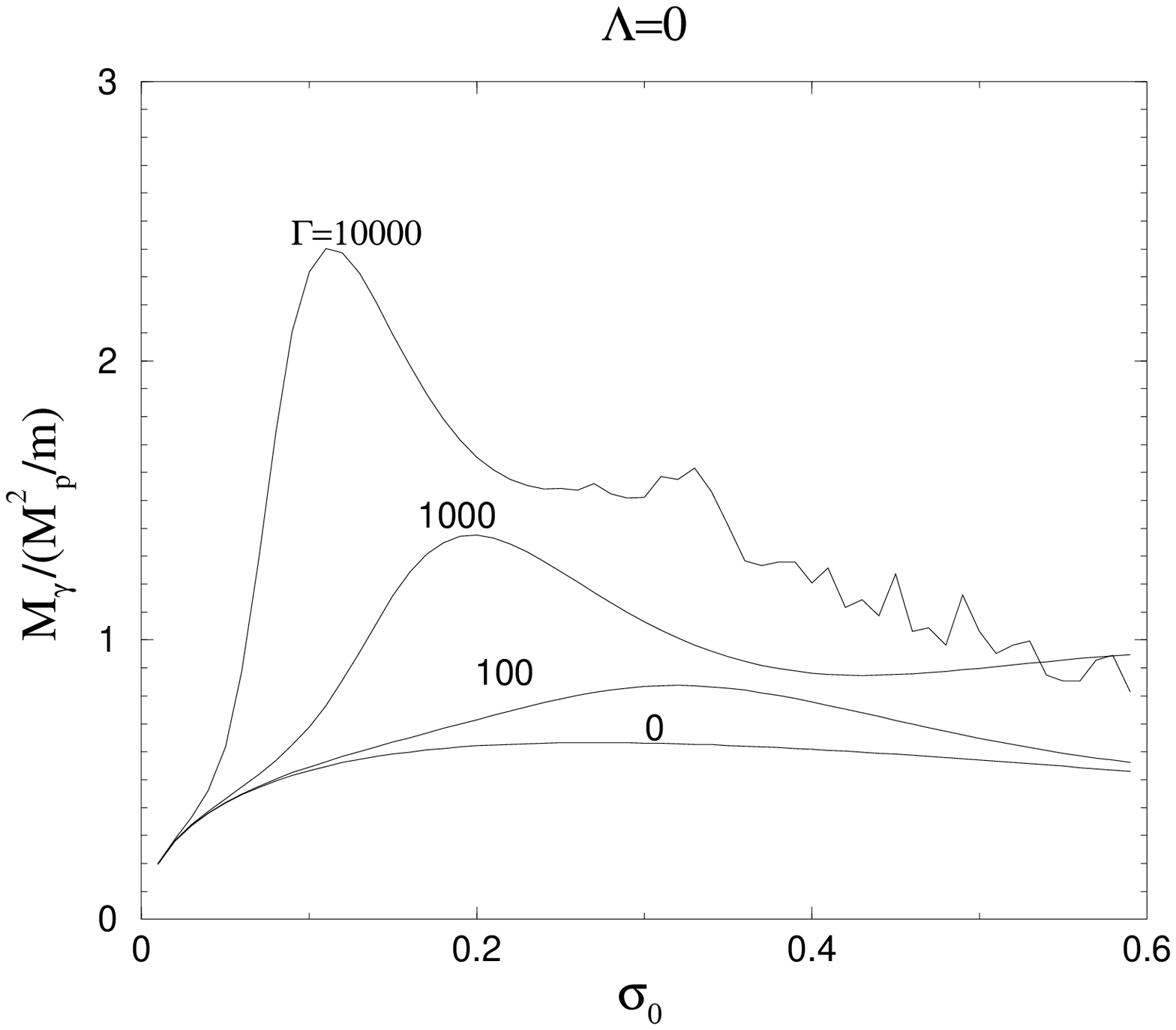,width=0.6\textwidth,angle=0}
    \caption{Boson star mass $M_{\gamma}$ as a function of $\sigma_0$
       when $\Lambda=0$}
    \label{fig:lambda0}
  \end{center}
\end{figure}

\begin{figure}[htbp]
  \begin{center}
    \leavevmode
    \epsfig{file=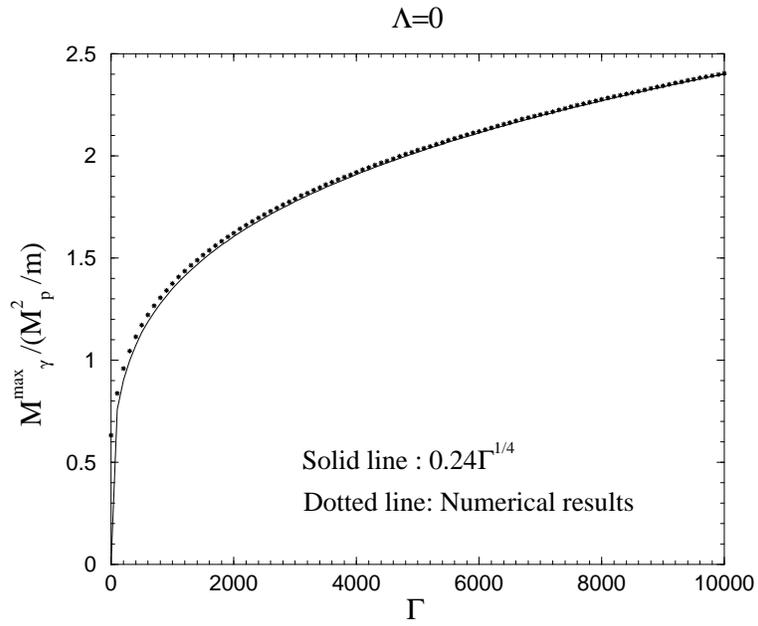,width=0.6\textwidth,angle=0}
    \caption{Maximum mass of boson star as a function of $\Gamma$.}
    \label{fig:maxmass}
  \end{center}
\end{figure}

First of all, we consider $\Lambda=0$ case. We can see in
Fig.\ref{fig:lambda0} that the boson star mass
increases with increasing $\Gamma$. Explicitly, from Fig.\ref{fig:maxmass}
plotted
the maximum mass of boson star as a function of $\Gamma$, we find out that
the maximum mass of boson star behaves as
\begin{equation}
  \label{maxgamma1}
  M^{max}_{\gamma} \approx 0.24\Gamma^{1/4}M^2_{p}/m
\approx 0.24 \tilde{\gamma}^{1/4}\left( \frac{m}{M_p} \right)^{1/2}M_{ch}.
\end{equation}
This result corresponds to the maximum mass in (\ref{masslamgam}) with
$\Lambda =0 $.

\begin{figure}[htbp]
  \begin{center}
    \leavevmode
    \epsfig{file=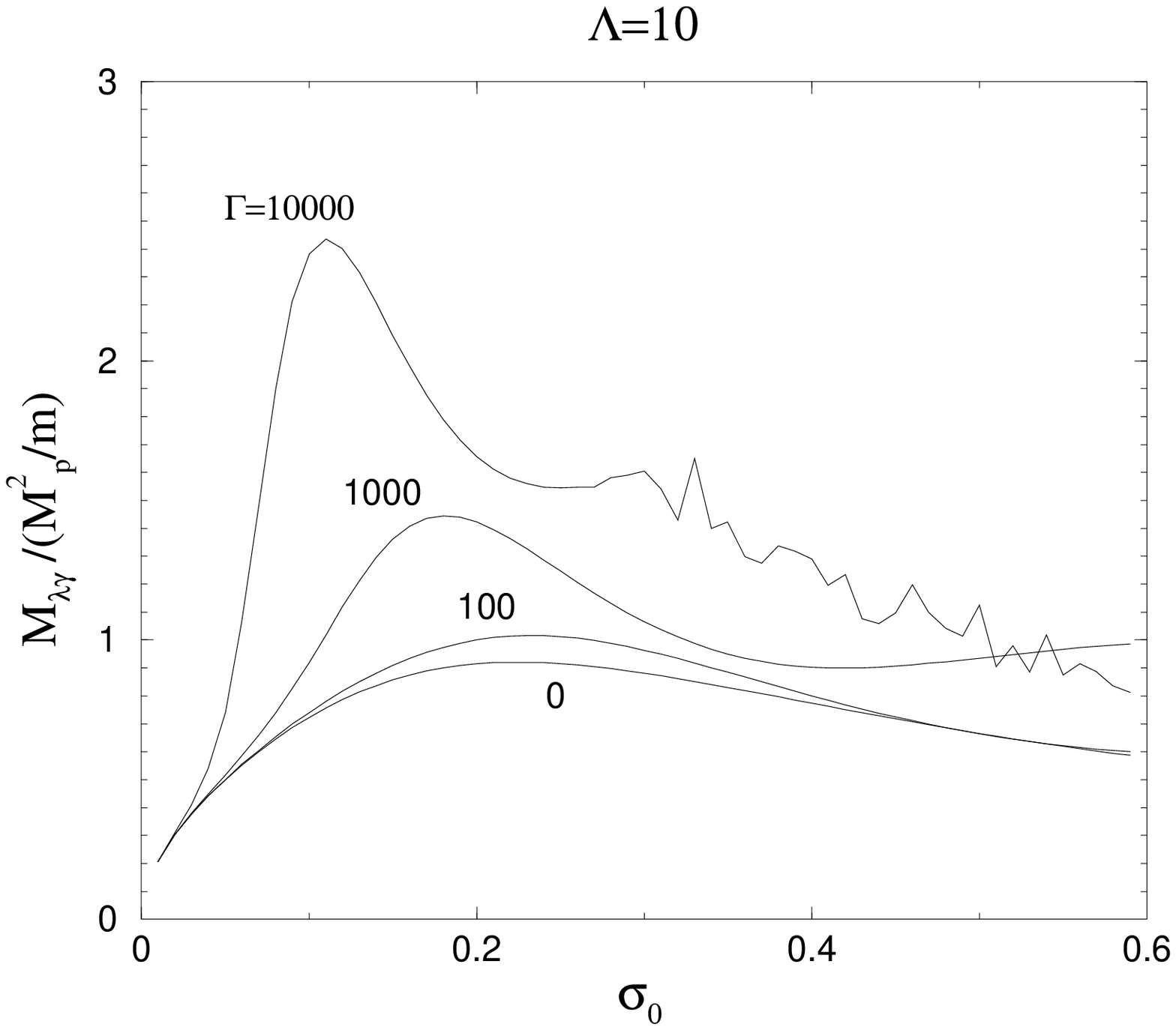,width=0.6\textwidth,angle=0}
    \caption{Boson star mass $M_{\lambda \gamma}$
       as a function of $\sigma_0$ when $\Lambda=10$}
    \label{fig:lambda10}
  \end{center}
\end{figure}

\begin{figure}[htbp]
  \begin{center}
    \leavevmode
    \epsfig{file=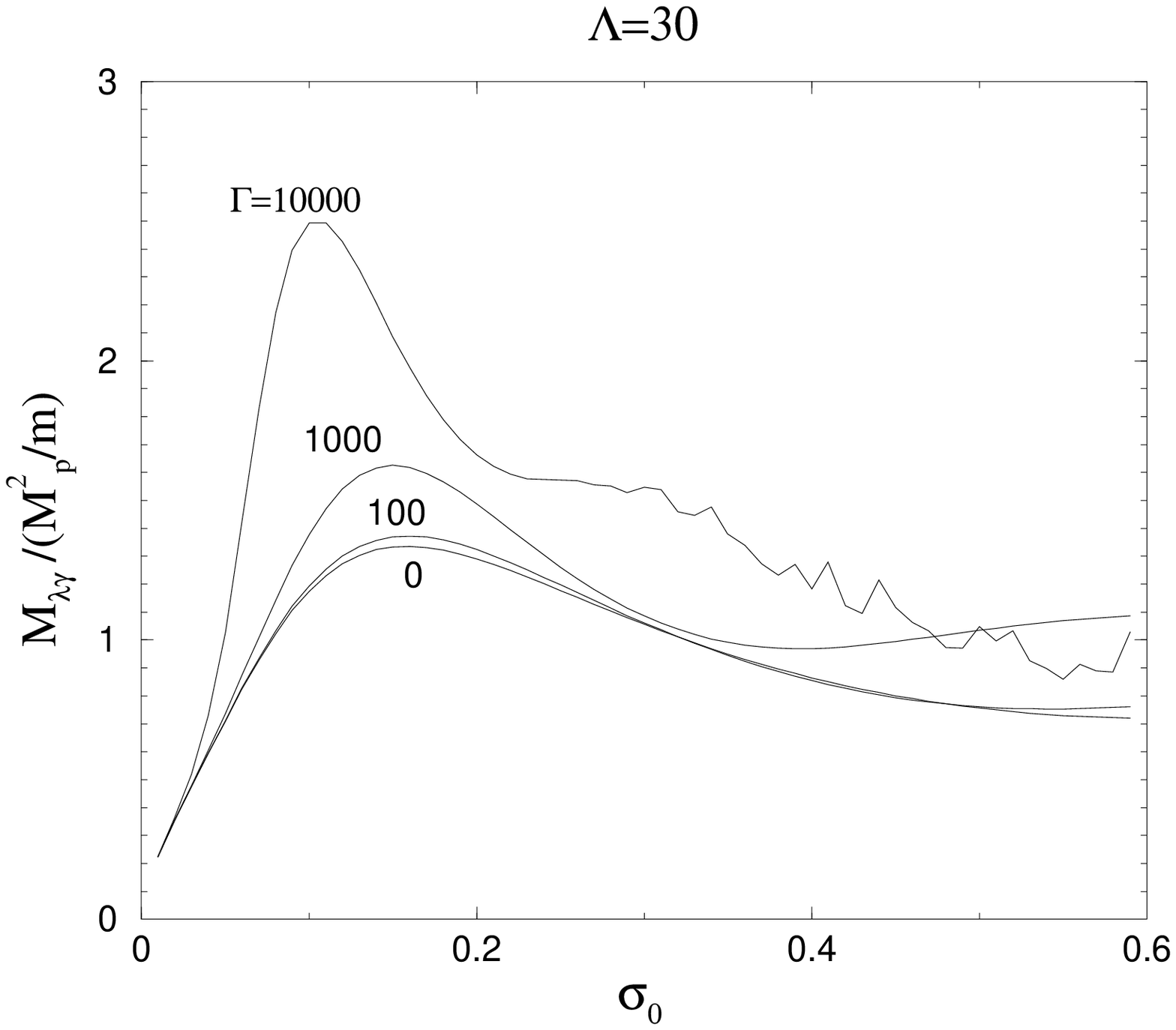,width=0.6\textwidth,angle=0}
    \caption{Boson star mass $M_{\lambda \gamma}$
as a function of $\sigma_0$ when $\Lambda=30$}
    \label{fig:lambda30}
  \end{center}
\end{figure}

\begin{figure}[htbp]
  \begin{center}
    \leavevmode
    \epsfig{file=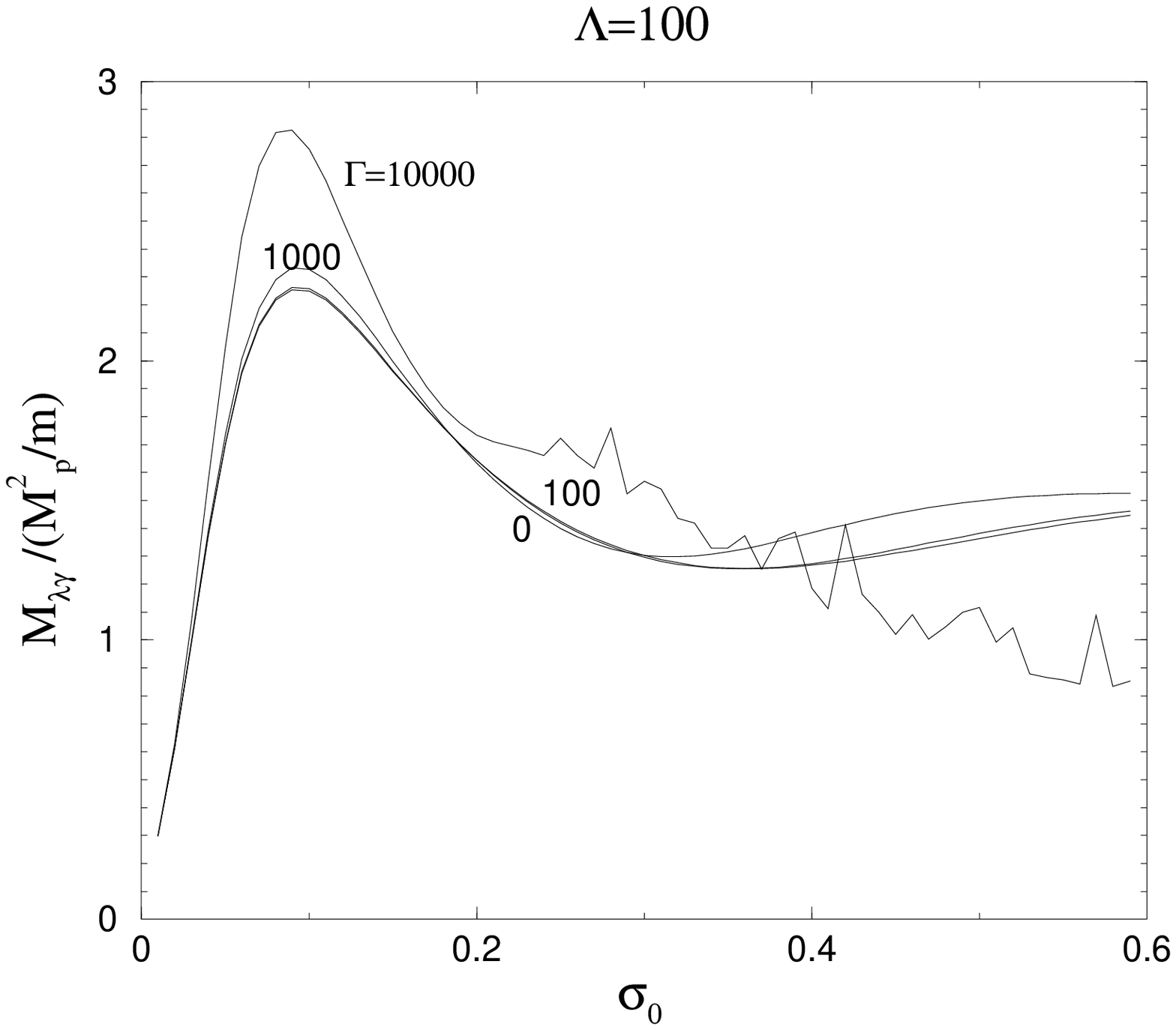,width=0.6\textwidth,angle=0}
    \caption{Boson star mass $M_{\lambda \gamma}$
as a function of $\sigma_0$ when $\Lambda=100$}
    \label{fig:lambda100}
  \end{center}
\end{figure}

\begin{figure}[htbp]
  \begin{center}
    \leavevmode
    \epsfig{file=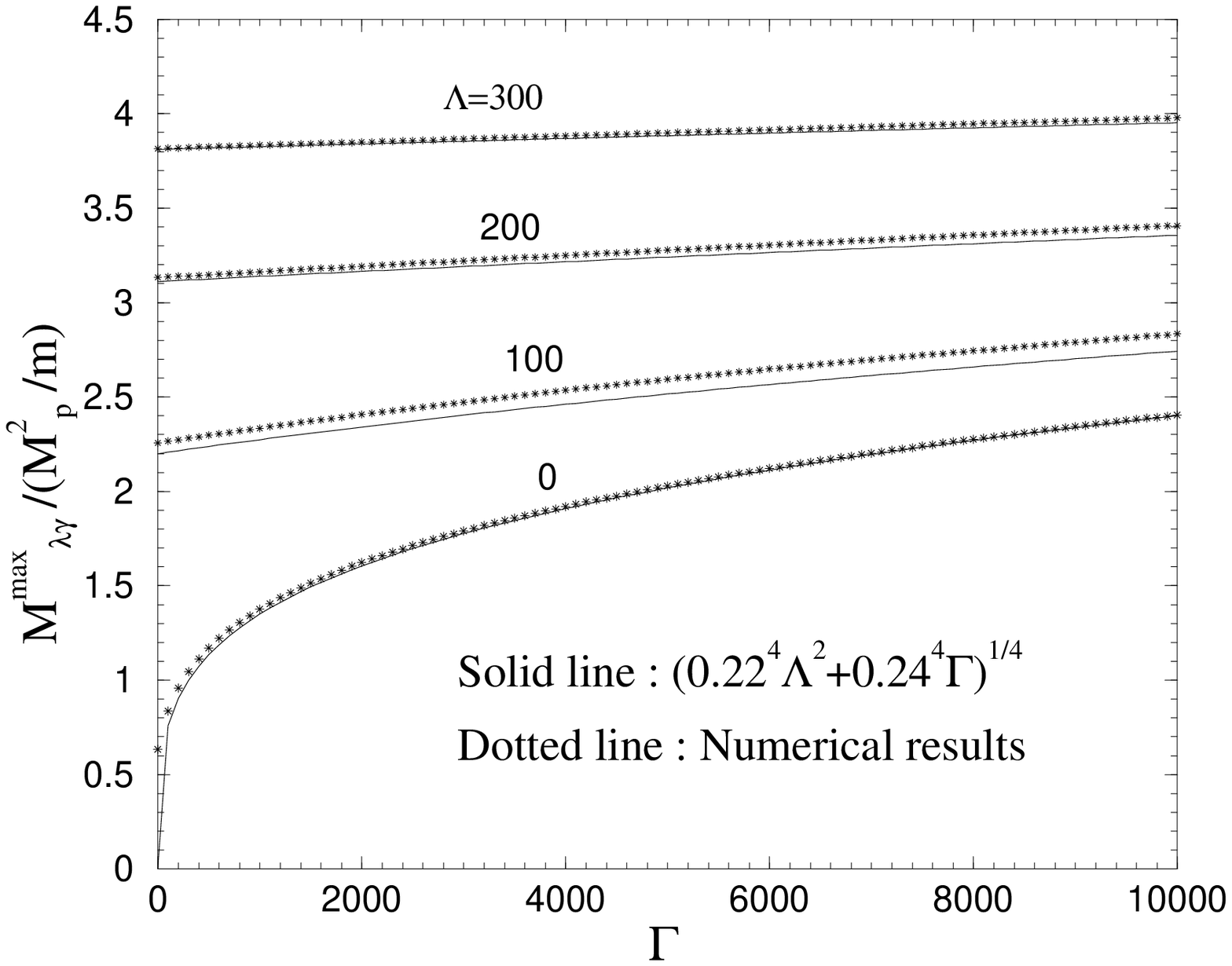,width=0.6\textwidth,angle=0}
    \caption{Maximum mass of boson star as a function of $\Gamma$ under $\Lambda=0, 100, 200, 300$.}
    \label{fig:maxcomp}
  \end{center}
\end{figure}

Let us now consider $\Lambda\ne 0$ case. Numerical results of
$\cal M\left(\infty\right )$ for
$\Lambda=10, 30, 100$ are shown in
Figs.\ref{fig:lambda10},\ref{fig:lambda30},\ref{fig:lambda100}.
We can see that $\Gamma$
dependence is not sensible for large value of $\Lambda$. 
From Fig.\ref{fig:maxcomp} in which we plot the maximum mass of boson star
as a function of
$\Gamma$ for several different $\Lambda$ values. 
It can be seen that dotted lines representing numerical results approach the
solid lines drawing $\left(0.22^4\Lambda^2+0.24^4\Gamma\right )^{1/4}$.
Thus, the maximum mass formed by the self-interacting scalar field is
given by
\begin{eqnarray}
  \label{maxlaga1}
  M^{max}_{\lambda \gamma}
&\approx &\left(0.22^4\Lambda^2+0.24^4\Gamma\right )^{1/4}
M^2_{p}/m
\nonumber \\
&\approx &\left(0.22^4\lambda^2+0.24^4 \tilde{\gamma}
\left(\frac{m}{M_p} \right)^2\right )^{1/4} M_{ch}.
\end{eqnarray}
This is equivalent to Eq.(\ref{masslamgam}) up to constant
factors.

Figs.1,3,4,5 show that the numerical noises become
noticeable when $\Gamma$ reaches about 10000.
However, the maximum masses of boson stars are
not affected by this part of the numerical error.

\section{Summary and Discussions}
Boson stars are only prevented from gravitational collapse by the 
Heisenberg uncertainty principle whose characteristic length scale is
enormously small than that of the Pauli exclusion principle. Thus, the
radius and mass of a boson star are much less than those of a fermionic
star. However, Colpi, Shapiro, and Wasserman \cite{colpi} have shown that
introducing the self-interacting scalar fields, the maximum mass
of a stable boson star may be comparable to the Chandrasekhar mass.

In this paper, we have made analytic derivation for the maximum mass and
have shown that it is equivalent to their numerical result. In addition,
it has been shown that the radius and density of the boson star formed by
the self-interacting scalar fields become larger and dilute compared with
those
 of the boson star formed by the scalar fields without self-interaction.
The origin of this drastic change is that the introduced self-interaction
term in matter Lagrangian generates a repulsive force. Thus, the
characteristic length scale of the theory becomes large, and the boson star
becomes large and massive. We also generalized our analytic derivation for
the boson star formed by scalar fields with any higher order
self-interactions. It has been shown that the contribution of the
higher order self-interaction terms to the maximum mass decreases as
$(m/M_p)^2$ power as expected in the viewpoint of the field theory.
Our analytic calculations have been confirmed in numerical analysis in which
we considered the scalar field with the quartic and the sixth order
self-interactions.

Recently, rotating boson stars have been also considered \cite{schunck}.
This generalization is likely to be natural and it would be interesting
to extend our analytic calculation to the rotating boson stars.

\section*{Acknowledgments}
S. K. and B.-H. L. were supported in part by Basic Science 
Research Institute Program, Ministry of Education, Project
No. BSRI-98-2414.



\newpage

\begin{thebibliography}{99}
\bibitem
{kolb} E.W. Kolb and M.S. Turner, {\it The Early Universe}
(Addison-Wesley, 1990).
\bibitem
{linde} A. Linde, {\it Particle Physics and Cosmology}
(Gordon and Breach, 1990).
\bibitem
{kaup} D.J. Kaup, Phys. Rev. {\bf 172} (1968) 1331.
\bibitem
{ruffini} R. Ruffini and S. Bonazzola, Phys. Rev. {\bf 187} (1969) 1767.
\bibitem
{colpi} M. Colpi, S.L. Shapiro and I. Wasserman, Phys. Rev. Lett. 
{\bf 57} (1986) 2485.
\bibitem
{bij} J.J. van der Bij and M. Gleiser, Phys. Lett. B{\bf 194} (1987) 482.
\bibitem
{jetzer} P. Jetzer and J.J. van der Bij, Phys. Lett. B{\bf 227} (1989) 341.
\bibitem
{henriques} A.B. Henriques, A.R. Liddle and R.G. Moorhouse,
Phys. Lett. B{\bf 233} (1989) 99; {\it ibid}, Nucl. Phys. B{\bf 337}
(1990) 737.
\bibitem
{gleiser} T.D. Lee and Y. Pang, Nucl. Phys. B{\bf 315} (1989) 477;
M. Gleiser and R. Watkins, Nucl. Phys. B{\bf 139} (1989) 733;
P. Jetzer, Nucl. Phys. B{\bf 316} (1989) 411; {\it ibid} Phys.
Lett. B{\bf 222} (1989) 447;
M. Gleiser, Phys. Rev. D{\bf 38}, (1988) 2376 [E{\bf 39} (1989)
1257].
\bibitem
{kusmartsev} F.V. Kusmartsev, E.W. Mielke and F.E. Schunck,
Phys. Rev. D{\bf 43} 3895.
\bibitem
{peebles} P.J.E. Peebles, {\it The Large Scale Structure of the Universe}
(Princeton University Press, 1980).
\bibitem
{liddle} A.R. Liddle and M.S. Madsen, Int. J. Mod. Phys. D{\bf 1} (1992) 101.
\bibitem
{mielke} E.W. Mielke and F.E. Schunck, {\it Boson Stars: Early History and
Recent Prospects}, gr-qc/9801063 (Proc. 8th M. Grossmann Meeting, T. Piran
(ed.), World Scientific, Singapore, 1998)
\bibitem
{tkachev} I.I. Tkachev, Sov. Astron. Lett. {\bf 12} (1986) 305.
\bibitem
{mielke2} E.W. Mielke and R. Scherzer, Phys. Rev. D{\bf 24} (1981) 2111.
\bibitem
{benitez} J. Ben\'{\i}tez, A. Mac\'{\i}as, E.W. Mielke, O. Obreg\'{\o}n
and V.M. Villanueva, Int. J. Mod. Phys. {\bf 12} (1997) 2835.
\bibitem
{schunck} R. Ferrell and M. Gleiser, Phys. Rev. D{\bf 40} (1989) 2524;
F.E. Schunck and E.W. Mielke, in {\it Relativity and Scientific Computing},
edited by F.W. Hehl, R.A. Puntigam and H. Ruder (Springer, Berlin, 1996),
pp. 138-151; S. Yosida and Y. Eriguchi, Phys. Rev. D{\bf 56} (1997) 762.
\end{thebibliography}
\end{document}